\documentclass[journal]{IEEEtran}
\usepackage{cite}
\usepackage{amsmath,amssymb,amsfonts}
\usepackage{algorithm}  
\usepackage{algpseudocode}  
\usepackage{algpseudocode}
\usepackage{graphicx}
\usepackage{epstopdf}
\usepackage{textcomp}
\usepackage{xcolor}
\usepackage{bm}
\usepackage{setspace}
\usepackage{subfigure}

\begin{document}
\title{Transmissive RIS Enabled Transceiver Systems: Architecture, Design Issues and Opportunities}
\author{Zhendong~Li,~Wen~Chen,~Qingqing~Wu,~Ziwei~Liu,~Chong~He,~Xudong~Bai,~and~Jun~Li
\thanks{This work is supported in part by National key project 2020YFB1807700, NSFC 62071296, and Shanghai Kewei 22JC1404000, in part by NSFC 62371289, NSFC 62331022, and Xiaomi Young Scholar Program, in part by the Key Technologies R\&D Program of Jiangsu (Prospective and Key Technologies for Industry) under Grants BE2023022 and BE2023022-2, in part by Shanghai Aerospace Science and Technology Innovation Foundation under Grant F-202405-0018, Basic Research Programs of Taicang under Grant TC2022JC16. Z. Li is with the department of Electronics Engineering, Shanghai Jiao Tong University and the School of Information and Communication Engineering, Xi'an Jiaotong University, Xi'an 710049, China (email: lizhendong@xjtu.edu.cn). W. Chen ({\em corresponding author}), Q. Wu, Z. Liu and C. He are with the Department of Electronic Engineering, Shanghai Jiao Tong University, Shanghai 200240, China (e-mail: wenchen@sjtu.edu.cn; qingqingwu@sjtu.edu.cn; ziweiliu@sjtu.edu.cn; hechong@sjtu.edu.cn). X. Bai is with the School of Microelectronics, Northwestern Polytechnical University, Taicang 215400, Suzhou, China (e-mail: baixudong@nwpu.edu.cn). J. Li is with the School of Information Science and Engineering, Southeast University, Nanjing, 210096, China (e-mail: jleesr80@gmail.com).}
}

%

\maketitle

\begin{abstract}
Reconfigurable intelligent surface (RIS) is anticipated to augment the performance of beyond fifth-generation (B5G) and sixth-generation (6G) networks by intelligently manipulating the state of its components. Rather than employing reflective RIS for aided communications, this paper proposes an innovative transmissive RIS-enabled transceiver (TRTC) architecture that can accomplish the functions of traditional multi-antenna systems in a cost-effective and energy-efficient manner. First, the proposed network architecture and its corresponding transmission scheme are elaborated from the perspectives of downlink (DL) and uplink (UL) transmissions. Then, we illustrate several significant advantages and differences of TRTC compared to other multi-antenna systems. Furthermore, the downlink modulation and extraction principle based on time-modulation array (TMA) is introduced in detail to tackle the multi-stream communications. Moreover, a near-far field channel model appropriate for this architecture is proposed. Based on the channel model, we summarize some state-of-the-art channel estimation schemes, and the channel estimation scheme of TRTC is also provided. Considering the optimization for DL and UL communications, we present numerical simulations that confirm the superiority of the proposed optimization algorithm. Lastly, numerous prospective research avenues for TRTC systems are delineated to inspire further exploration.
\end{abstract}

\section{Introduction}
\IEEEPARstart{T}{he} evolution of mobile communication has led to an increase in both the power consumption and cost of conventional base station (BS). Specifically, in fifth-generation (5G) networks, the signal propagation loss is exacerbated due to higher frequency band operating characteristics compared to fourth-generation (4G) networks. Consequently, achieving comparable coverage capabilities requires several times more BSs. Furthermore, the extensive deployment of multi-input multi-output (MIMO) mandates a proportional augmentation in the quantity of signal processing modules and active radio frequency (RF) chains to be deployed on the BS. This inevitably results in a higher power consumption of the 5G networks compared to their 4G counterparts \cite{9113273}. Thus, one of the prevailing concerns with current 5G BS is their high power consumption, design cost, and deployment cost compared to 4G BS. Hence, the imperative is to devise an innovative transceiver architecture suited for the future beyond fifth-generation (B5G) and sixth-generation (6G) networks, capable of attaining reduced power consumption and cost.

Recently, an innovative device with considerable potential has been proposed and named the reconfigurable intelligent surface (RIS). This planar array consists of numerous cost-effective and passive elements, the amplitudes and phase shifts of which can be independently controlled through the intelligent controller equipped with the RIS, allowing directional beams to be generated. Since RIS has no signal processing capability, it is less affected by interference and noise, which is beneficial for signal transmission. Moreover, the passive nature of the RIS makes it simpler in structure and requires fewer complex components, resulting in reduced system power consumption and cost. The distinctive characteristics of RIS position it as a promising candidate for the next-generation communication networks.

The RIS is categorized into reflective, transmissive, stacked and hybrid, with recent research focusing on RIS-assisted communication using these modes \cite{9531372,9509394,10158690,9200683}. By independently altering the phase shift and amplitude of passive element, RIS can effectively reconstruct the wireless channel, leading to improved signal strength of desired transmissions and reduced interference from unwanted signals. The excellent portability of RIS makes it an ideal candidate for auxiliary communication in diverse scenarios without necessitating any alterations to the existing network infrastructure, thereby boosting quality-of-service (QoS) of users.

The use of RIS is not limited to auxiliary communication, as it can also be implemented in both transmitters and receivers, thereby opening up two distinct research lines \cite{9133266,10177872}. It is worth noting that compared with conventional multi-antenna systems, since RIS-enabled transceivers do not need to be equipped with a large number of RF chains, the structure is simpler and power consumption and cost are also greatly reduced. In \cite{9133266}, authors proposed a reflective RIS transmitter architecture and analyzed its performance. Compared with the reflective RIS transmitter, the TRTC is free from feed blockage and echo self-interference problems. Therefore, it can be designed to be more efficient \cite{bai2020high,7448838}, which aligns well with the demands of future networks, as it facilitates a transceiver solution with significantly lower power consumption and reduced cost. Thus, TRTC is very promising in future communication networks.
\begin{figure*} [htbp]
	\centering
	\includegraphics[scale=0.5]{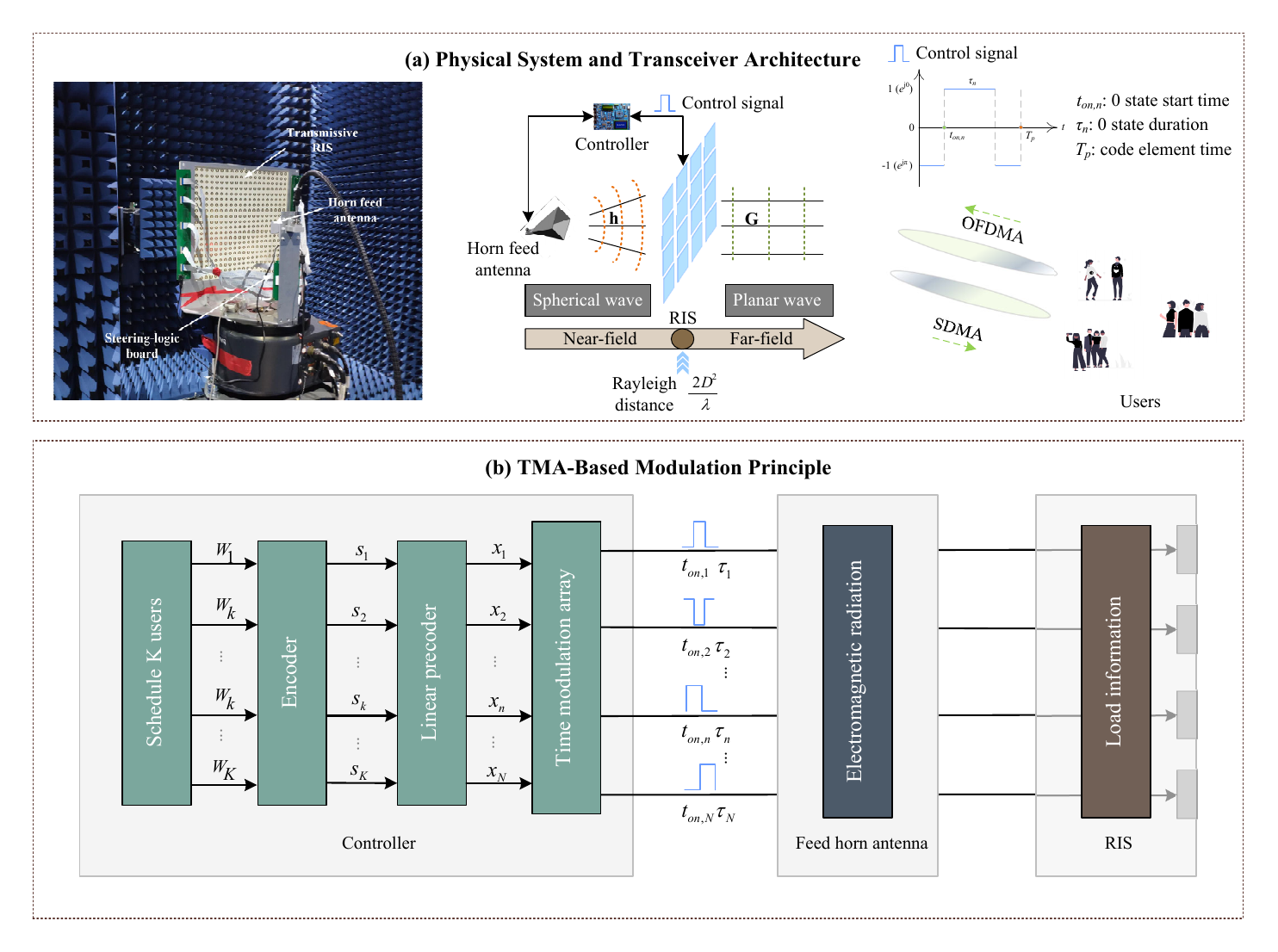}
	\caption{TRTC architecture diagram.}
\end{figure*}

While the TRTC system offers numerous benefits, it comprises a passive transmissive RIS and a single antenna horn feed, distinguishing it considerably from the conventional multi-antenna system that is composed of active components. At present, the investigation into the TRTC system remains in its early phases, and it is imperative to delineate its framework and mechanism. This rationale motivates us to present an overview of the TRTC system. In Section II, we elaborate the network architecture and transmission scheme. In Section III, the advantages and differences of TRTC system are introduced. In Section IV, we present the time-modulation array (TMA)-based downlink modulation and exaction principles for realizing multi-stream communications under this architecture. Section V of the paper presents the corresponding far-near field channel model along with some state-of-the-art channel estimation schemes. The downlink (DL) transmitter and uplink (UL) communication optimization are also given in Section VI, and some corresponding simulation results of two study cases verify the effectiveness of the proposed algorithm in TRTC system. Simultaneously, Section VII outlines several potential research directions and opportunities for future investigations. Ultimately, Section VIII concludes this paper.

\section{TRTC Architecture and Transmission Scheme}
The physical diagram and TRTC architecture are depicted in detail in Fig. 1(a), which is comprised of a single feed antenna, $N$ transmissive elements of RIS with an intelligent controller, and $K$ users. Fig. 1(b) shows the DL TMA modulation principle. Specifically, by controlling the switch period and duty cycle, TMA can generate different time-domain signals. These signals will exhibit specific spectral characteristics in the frequency domain, including the main beam direction and sidelobe levels. This method can be effectively applied to generate control signals for the passive RIS units in TRTC architecture. Besides, the materials used to make RIS in the TRTC architecture can be divided into PIN diodes, varactor diodes, graphene, and liquid crystal, etc. In DL, the horn feed antenna emits a single-frequency electromagnetic (EM) wave that does not convey information. Instead, all modulation and signal processing occur in the controller. Conversely, in UL, the receiving antenna captures signals from the users. The transmissive coefficient of the RIS can be mathematically represented as ${f_{{n}}} = {\beta _{{n}}}{e^{j{\theta _{{n}}}}}$, where ${\beta _{{n}}}$ and ${\theta _{{n}}}$ denote the amplitude and phase shift of the $n$-th element, respectively, with ${\beta _{{n}}} \in \left[ {0,1} \right]$ and ${\theta _{{n}}} \in \left[ {0,2\pi } \right)$. Moreover, the intelligent controller, integrated with the RIS, can dynamically adjust the amplitude and phase shift of the RIS transmissive element. This capability facilitates modulation and beamforming of the DL signal and enhances the UL incident signal. Considering the unique transceiver structure and the multi-user communication scenario, the signal is modulated and the multi-user beamforming scheme is designed in the controller for the DL phase, where space division multiple access (SDMA) is implemented. In the UL phase, the single feed antenna is responsible for receiving signals, and orthogonal frequency division multiple access (OFDMA) is employed to serve multiple users. It should be noted that the large-scale array of RIS can facilitate space division multiplexing, so we adopt SDMA in DL, but this architecture can also employ other access methods in DL, such as OFDMA and non-orthogonal multiple access (NOMA), etc. The impact on the system is similar to the effect these access methods have on the performance of traditional multi-antenna systems. The detailed schemes for DL and UL transmissions are outlined as follows:

\emph{DL Transmission:} The DL communications involve joint consideration of modulation signals and beamforming design by the intelligent controller equipped with RIS. The controller employs a TMA scheme to generate corresponding control signals. The essence of the TMA scheme is to modulate information onto harmonics. The phase shift of the RIS transmissive element is then changed through the control line, enabling the information to be loaded into the EM wave emitted by the feed horn antenna and transmitted to the users. Upon reception of the signal, different users perform operations such as harmonic extraction, symbol recovery, and demodulation on the received signal to obtain the desired information. This approach facilitates the realization of DL multi-user SDMA \cite{9570775}.

\emph{UL Transmission:} In the case of UL transmission, given the limitation of the single antenna equipped in the receiving horn, multi-user access is achieved through the utilization of OFDMA. Initially, on each subchannel, modulation signals are transmitted by the user. The uplink signal is initially received by the RIS, which amplifies and transmits it to the receiving antenna. Subsequently, the received signal is forwarded to the intelligent controller for additional processing, encompassing demodulation and decoding, etc. Notably, the enhancement of UL transmission performance is attained by optimizing the coefficients of RIS transmissive elements, including both magnitude and phase shift. As such, UL multi-user OFDMA is also realized.

\section{Comparing TRTC with Other Systems: Advantages and Differences}
This section will expound on the advantages and differences of TRTC in comparison to conventional multi-antenna transceivers and reflective RIS enabled transceivers.

\subsection{TRTC vs. Conventional Multi-Antenna Transceiver}
The TRTC system comprises a horn feed antenna, an intelligent controller, and a passive transmissive RIS. Compared to conventional multi-antenna transceivers, TRTC's structure is relatively uncomplicated, and it utilizes time-modulation array (TMA) for downlink multi-stream transmission. The advantages and differences of TRTC are as follows:

\emph{Relatively Low Power Consumption and Cost:} Unlike conventional multi-antenna systems that require multiple RF chains, the proposed TRTC utilizes only one RF chain for realizing multi-stream communications based on TMA, resulting in lower power consumption and cost. 
	
\emph{Practical Impact of TMA:} Diverging from conventional multi-antenna systems, the proposed TRTC architecture considers the transmission power constraints of each individual element. This is due to TMA employs non-linear modulation on harmonics. Notably, not all of the power received by the transmissive RIS element from the horn antenna and sent to the user carries information. As such, there exists a constraint on harmonic utilization, whereby the transmission power of each RIS element is limited to its maximum available useful power, i.e., the power required to carry information.
	
\emph{Relatively Simple Structure:} The conventional multi-antenna system typically has a complex signal processing module. TRTC, in contrast to conventional multi-antenna systems, requires only the generation of a control signal in the controller through TMA, which corresponds to the RIS element, to control its state. Additionally, the need for an up-conversion process is eliminated, simplifying the overall system design.
\begin{figure}
	\label{PAAbefore}
	\centering
	\subfigure[Magnitude for $\pi$-state.]{
		\begin{minipage}{8cm} 
			\includegraphics[width=\textwidth]{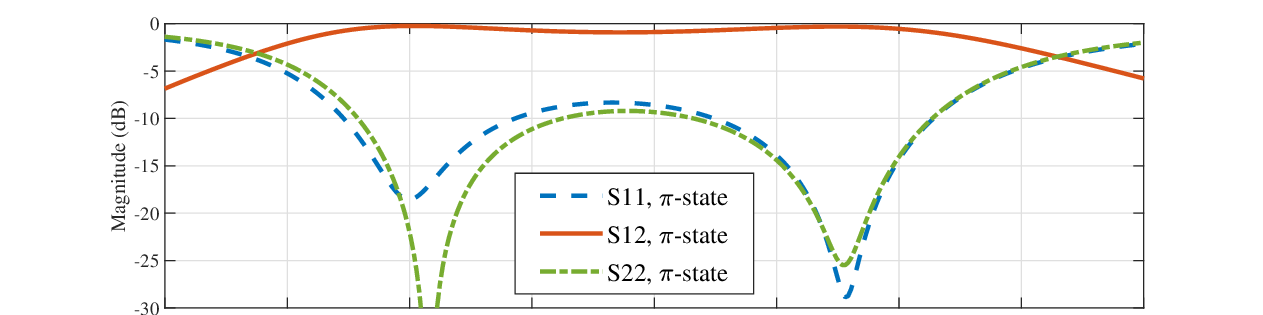} \\
		\end{minipage}
	}
	\subfigure[Magnitude for 0-state.]{
		\begin{minipage}{8cm}
			\includegraphics[width=\textwidth]{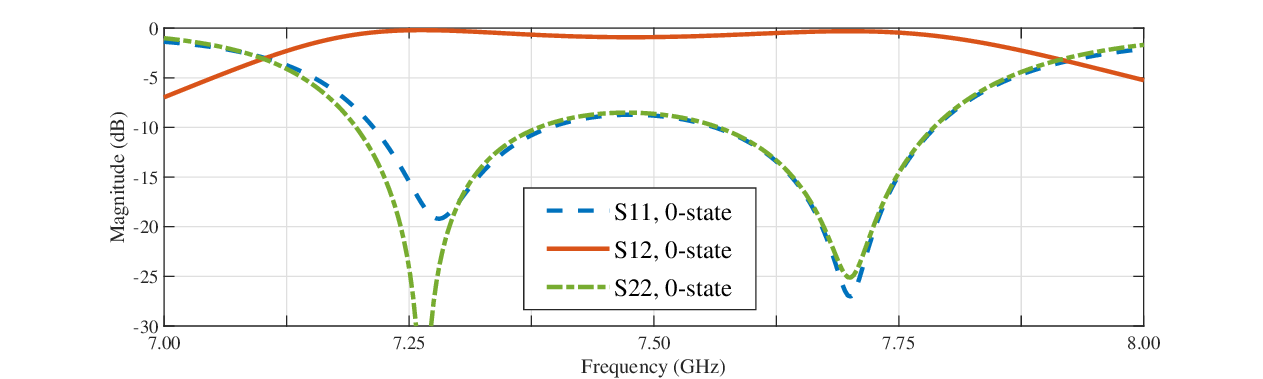} \\
			
		\end{minipage}
	}
	\caption{RIS coefficients in different phase shifts state.} 
\end{figure}
\subsection{TRTC vs. Reflective RIS Enabled Transceiver}
The TRTC system and the reflective RIS enabled transceiver system have different structures, resulting in natural advantages for the TRTC system, including:

\emph{Feed Blockage Free:} Feed blockage refers to the phenomenon that the feed horn antenna blocks the transmitted EM waves. In the architecture of a reflective RIS enabled transceiver, both the user and horn feed antenna are positioned on the same side of the RIS, leading to feed blockage issues for users located on that side. One solution to this problem is to reposition the horn antenna, but this would result in reduced transceiver efficiency. In contrast, the transmissive RIS transceiver architecture does not suffer from this issue, as the user and feed source are positioned on opposite sides of the RIS, which can significantly enhance the aperture efficiency of the transceiver.
	
\emph{Self-Interference Free:} Self-interference refers to the interference that the echo brings to the incident wave. In the reflective RIS transceiver, the feed source's incident EM wave and the RIS's reflected EM wave are on the same side, which inevitably causes echo interference issues. In contrast, TRTC eliminates the problem of self-interference by using transmissive RIS. In this setup, the incident and the transmitted EM wave are located on different sides of the RIS, which provides natural isolation and prevents interference between them. This renders TRTC a highly promising solution for future wireless communications, particularly in environments where the levels of interference and system complexity are anticipated to be elevated.
\begin{table*}[htbp]
	\centering
	\caption{The advantages and differences of TRTC vs. other system.}
	\renewcommand\arraystretch{1.0}
	\begin{tabular}{|c|c|c|c|}
		\hline &Conventional Multi-Antenna Transceiver&TRTC&Reflective RIS Enabled Transceiver\\
		\hline Power Consumption&High&Low&Low\\
		\hline Cost&High&Low&Low\\
		\hline Structure&Complicated&Simple&Simple\\
		\hline TMA&No&Yes&/\\
		\hline Feed Blockage&/&No&Yes\\
		\hline Echo Interference&/&No&Yes\\
		\hline Radiation Efficiency&/&High&Low\\
		\hline Working Bandwidth&/&Wide&Narrow\\
		\hline
	\end{tabular}
\end{table*}
	
\emph{Radiation Efficiency and Working Bandwidth Enhancement:} Leveraging the advantages offered by the aforementioned physical structure, the TRTC can attain higher aperture efficiency and expanded operating bandwidth \cite{bai2020high}. For the proposed architecture, Fig. 2(a) and Fig. 2(b) depict the magnitudes of the simulated scattering coefficients for a 1-bit transmitted RIS under various phase states. Here, S11 and S22 denote the scenarios involving reflective EM waves	, whereas S12 corresponds to the case of transmissive EM waves. It is noticeable that within the frequency range of 7.12 to 7.85 GHz, the transmissive coefficient consistently stays above -2 dB, while the reflective coefficient maintains a level below -10 dB, in both 0-state and $\pi$-state configurations. Consequently, the TRTC system can achieve enhanced radiation efficiency and expanded operating bandwidth with lower power consumption and cost.

Hence, Table I can be utilized to provide a comprehensive summary of the distinctive benefits and contrasts of the TRTC system when compared to other multi-antenna systems.

\begin{figure}
	\centerline{\includegraphics[width=9cm]{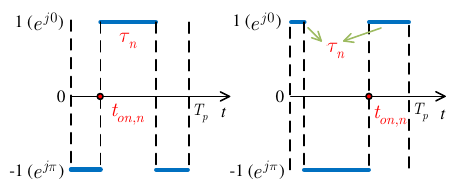}}
	\caption{The waveforms of control signal with two degrees of control freedom ${t_{on,n}}$ and ${\tau _n}$.}
\end{figure}
\section{TMA-Enabled DL Modulation and Extraction Techniques for DL Multi-Stream Communications}
This section focuses on presenting the DL modulation principle of the TRTC system based on TMA and extraction technique for realizing multi-stream communications. The DL transmission scheme in the proposed architecture differs significantly from conventional architecture. It is worth noting that the TRTC lacks complex RF chains and signal processing capabilities, which makes the direct application of conventional transmission schemes impossible. Therefore, we adopt TMA for the DL multi-stream communication of the system.

\emph{Generation of Multi-Stream Control Signals:} Let $s_k$ represent the modulated signal to be sent by the transceiver to user $k$, which can be a modulated signal of any order, then the signal transmitted by the transceiver to all users can be represented by ${\bf{s}} = {\left[ {{s_1},...,{s_K}} \right]^T} \in {\mathbb{C}^{K \times 1}}$. In addition, let ${{\bf{f}}_k} = {\left[ {{f_{1,k}},...,{f_{N,k}}} \right]^T} \in {\mathbb{C}^{N \times 1}}$ denote beamforming vector for the $k$-th user. We denote the beamforming matrix for all users as ${\bf{F}} = \left[ {{{\bf{f}}_1},...,{{\bf{f}}_K}} \right] \in {\mathbb{C}^{N \times K}}$. Therefore, the transmission signal ${\bf{x}} \in {\mathbb{C}^{N \times 1}}$ of the novel transceiver can be expressed as ${\bf{x}} = {\bf{Fs}}$, which is a vector of dimension $n$, and the transmission signal $x_n$ of each element contains all the signal components needed by the users. The synthesized modulation signal symbol, formed by superimposing these components, can be expressed as ${A_n}{e^{j{\phi _n}}}$, where $A_n$ and $\phi_n$ represent the amplitude and phase shift, respectively. These parameters can be mapped onto control signal waveforms in binary states (0 and 1) using TMA \cite{9576618}, which in turn can control the state of RIS transmissive elements. Thus, this enables the transmission of information. Fig. 3 illustrates that the adoption of TMA provides the control signal waveforms with two degrees of control freedom, namely, the phase start time ${t_{on,n}}$ and conduction duration ${\tau _n}$. Through the expansion of the Fourier series for the control symbol, the amplitude and phase shift on the $l$-order harmonic can be obtained. By configuring these two degrees of freedom, the amplitude and phase shift of $l$-order harmonics can be independently adjusted. Therefore, TMA can modulate information onto harmonics and is essentially a non-linear modulation scheme. Then, the information can be loaded into the carrier wave emitted by the feed horn antenna and transmitted to the user. Thus, any-order modulation signal and beamforming vector can be jointly transmitted by applying the TMA scheme. Additionally, the scheme is reconfigurable, and the control signal waveform can be adjusted based on different modulation and beamforming requirements.

\emph{Extraction of User Signals:} As TMA modulates the signal on the harmonic, it is necessary to extract the harmonic modulation signal for each user after receiving the signal. This can be achieved by calculating the harmonic components of the received signal through either single-point or two-point fast Fourier transformation. Once the composite modulated signal symbols are recovered, the superimposed signals of all users can be obtained, enabling the utilization of conventional communication system schemes for demodulation, decoding, and other processes.

\section{Far-Near Field Channel Model and Channel Estimation in TRTC}
\subsection{Far-Near Field Channel Model}
Fig. 1(a) illustrates that in wireless communication environments, the distance between the field point and the source allows for a distinction between near-field radiation and far-field radiation, leading to the development of corresponding channel models \cite{cui2021channel}. The distance metric is calculated using the Rayleigh distance, which is given by $\frac{2D^2}{\lambda}$, where $D$ and $\lambda$ denote the array aperture and wavelength, respectively. Specifically, when the distance between the field point and the wave source exceeds the Rayleigh distance, the wavefront in that region is approximately a uniform plane wave according to the wavefront hypothesis theory. Consequently, the field point's radiation is in the far-field region, allowing the channel to be modeled as a far-field. However, if the distance between the field point and the wave source is less than the Rayleigh distance, the wavefront at that moment becomes a spherical wave, and the field point's radiation is near-field. Hence, the channel is modeled as a near-field.

\emph{Far-Field RIS-User Channel Model:} As the physical distance between the user and the RIS is much larger than the Rayleigh distance, the field in which the user is situated falls under the far-field region. As a result, the RIS-user channel can be modeled as a far-field channel, where the wavefront is approximated as a uniform plane wave. In practice, the user-RIS channel is composed of both a line-of-sight (LoS) component and a non-line-of-sight (NLoS) component, which are superimposed. Consequently, the channel between the RIS and the user is characterized as a Rician fading channel. Setting the Rice factor to zero transitions the RIS-user channel into a Rayleigh fading channel. Furthermore, each element of the NLoS component is modeled as an independent and identically distributed circularly symmetric complex Gaussian (CSCG) random vector with zero mean and unit variance. For the LoS component, modeling can be undertaken using either of two approaches: the uniform linear array (ULA) or the uniform planar array (UPA).

\emph{Near-Field RIS-Horn Channel Model:} The channel from the RIS to the horn feed antenna can be referred to as the RIS-horn channel. Since the distance between the transmitter and the receiver is less than the Rayleigh distance, the EM radiation field in the surrounding space is considered to be in the near-field region, and the wavefront has a spherical shape. Additionally, there are no physical objects in the space between the RIS and the feed antenna that may cause EM wave propagation blockage or fading, thus allowing the use of a LoS channel model to represent the RIS-horn channel. According to the transmission mechanism, it needs to be focused on in the uplink design of the TRTC architecture.

\subsection{Channel Estimation}
Obtaining channel state information (CSI) is crucial for both RIS transmissive coefficient designs and communication optimizations in DL and UL. Channel estimation is particularly challenging for TRTC, as the transmissive element of the RIS can only passively transmit signals. Additionally, since users use SDMA and OFDMA in DL and UL, respectively, CSI of DL and UL cannot be obtained through the reciprocity of the channels. Consequently, under this framework, distinct channel estimation schemes for DL and UL must be developed.

\emph{DL Channel Estimation:} In DL transmission, the transmissive RIS transceiver generates an information-carrying control signal in the controller and loads it onto the EM waves emitted by the horn feed antenna. Therefore, to estimate the DL channel, we only need to focus on estimating the RIS-user channel. It is worth noting that the estimation of the RIS-user channel takes place at the user's end. To accomplish this, we first employ the Lloyd-based codebook design algorithm, which was proposed in \cite{6197256}, to synchronize the channel codebook offline at the TRTC and the users. Subsequently, a pilot sequence of a specific length is transmitted, which allows for the channel estimation to be obtained at the user end by utilizing either the least square (LS) criterion or the minimum mean square error (MMSE) criterion. Following this, when the users request communication, the index information is uploaded to the TRTC, which, in turn, obtains the CSI of the DL based on the received index and then proceeds with the RIS transmissive coefficient design.

\emph{UL Channel Estimation:} Regarding UL transmission, users convey their own information on each sub-channel, which is then received by the horn feed antenna after being transmitted by the RIS. Therefore, to estimate the UL channel, we must assess two channel components, namely the RIS-user channel and the RIS-horn channel. Moreover, channel estimation can only take place at the horn feed antenna due to the RIS transmissive element's lack of acceptance capability. As a result, the CSI of the UL can only be obtained via direct cascade channel estimation, while the CSI of the separable channel cannot be determined directly. It is important to highlight that the near-field RIS-horn channel can be either calculated or measured and is generally considered to be constant. In scenarios where the cascaded CSI and near-field CSI are known, RIS separable cascaded channel estimation methods can be employed to acquire the far-field CSI. We will now elaborate on the two channel estimation techniques, namely direct and separable cascaded channel estimation.

In the case of the direct cascaded channel, which involves the multiplication of two channel parts, existing literature typically suggested channel estimation protocols followed by various estimation algorithms \cite{8879620}. Common direct cascaded channel estimation algorithms include message passing (MP), channel correlation, optimization-based, decomposition, and interpolation recovery methods. It should be noted that the MP algorithm typically requires a sparse representation of the channel, while considering channel correlation can reduce pilot overhead.

Several studies proposed methods for separable cascaded channel estimation, including semi-passive RIS and fully passive RIS methods \cite{9361077}. In semi-passive RIS, a portion of active elements receive the transmitted signal from the user or transceiver, and estimate the path loss and angle of the RIS-horn and RIS-user channels. However, this method requires feedback to the transceiver and user for beamforming design, which can reduce efficiency and increase hardware costs. On the other hand, full-passive RIS methods can overcome these limitations. In fully passive RIS, the user transmits different pilot sequences to the transceiver through the RIS, and the cascade channel can be estimated through the analysis of the received signal. It should be emphasized that the near-field RIS-horn channel of the TRTC system can be obtained through measurement or calculation, and the cascaded channel can be obtained through the method discussed in the previous subsection.

\section{DL and UL Case Studies for TRTC System}
\subsection{Optimization of DL Communications}
Based on the DL transmission mechanism and CSI of RIS-user channel utilizing the channel estimation scheme discussed above, the DL communications can be optimized. In DL, the objective is to maximize the sum-rate by jointly optimizing the RIS transmissive coefficient and power allocation. However, due to the high coupling of the optimization variables, the formulated optimization problem is inherently non-convex. To tackle this challenge, we adopt the alternating optimization algorithm, which allows us to decouple the original problem into two tractable sub-problems. To address this, we employ difference-of-convex (DC) programming and successive convex approximation (SCA) techniques to transform the sub-problems into convex forms. Then, these sub-problems are subsequently solved using CVX toolbox. The optimization of sub-problems proceeds in an alternating manner until convergence is achieved. In simulation settings, we consider a system with $K=4$ users and $N=25$ RIS elements, where all user parameters are assumed to be the same. To evaluate the effectiveness of our proposed algorithm, we compare it with three benchmark algorithms: (1) Benchmark 1, which adopts the same RIS transmissive coefficient design as our proposed scheme and uses the equal allocation scheme for the power allocation of multiple users. (2) Benchmark 2, which utilizes the zero-forcing (ZF) beamforming scheme and equal allocation scheme for the power allocation of multiple users. (3) Benchmark 3, which adopts a random allocation approach to achieve beamforming and power allocation.

\subsection{Optimization of UL Communications}
Upon acquisition of the UL CSI through the aforementioned channel estimation scheme, the UL communications can be optimized. In this system, an optimization problem can be formulated to maximize the achievable system sum-rate. This is achieved by jointly optimizing the power allocation, subcarrier allocation, and the RIS transmissive coefficient. However, this optimization presents a significant challenge, as it constitutes a mixed-integer non-convex problem. To effectively address this, the alternating optimization algorithm can be employed, utilizing methods such as Lagrangian dual decomposition, DC, SCA, and penalty function. The system setup for UL consists of a 3D coordinate system with $K=5$ users and $N=25$ transmissive elements, with all user parameters assumed to be the same. To evaluate the effectiveness of the proposed algorithm, we compare it with the following benchmark algorithms: (1) Benchmark 1, which involves optimizing the three variables sequentially and without alternating optimization. (2) Benchmark 2, which employs Lagrangian dual decomposition method to optimize the subcarrier and power allocation, while the RIS transmissive coefficient is generated randomly. (3) Benchmark 3, where a random generation strategy is used to optimize all three sets of variables.
\subsection{Simulations of Two Case Studies}
Fig. 4 presents the results of numerical simulations conducted on the DL and UL scenarios using the aforementioned optimization algorithms. Specifically, Fig. 4(a) illustrates the changes in the DL system sum-rate as a function of the number of RIS transmissive elements for different benchmark algorithms. It is apparent that the system sum-rate exhibits a consistent trend across all benchmark algorithms, whereby an increase in the number of RIS elements leads to a corresponding increase in the system sum-rate. This can be attributed to the fact that a larger number of RIS transmissive elements yields higher spatial diversity gain. Meanwhile, the inclusion of more RIS transmissive elements results in an increase in the number of signal scattering paths, which enhances the signal received at the user compared to the previous setup. The superior performance of the proposed algorithm can be observed in comparison to the other benchmark algorithms for the same number of RIS transmissive elements, which signifies the effectiveness of the proposed algorithm in achieving optimal system performance. Additionally, Fig. 4(b) demonstrates the variation in UL system performance with the number of RIS transmissive elements for different benchmark algorithms. It clearly illustrates that the system sum-rate increases with the number of RIS transmissive elements under different benchmark algorithms. Furthermore, the superiority of the proposed algorithm can be attributed to the fact that the optimal RIS transmissive element coefficients provide substantial signal gain, which effectively boosts the UL transmission signal at the user end. Hence, an increase in the number of RIS transmissive elements translates to a higher overall gain, which ultimately boosts the system sum-rate. This finding offers valuable insights for TRTC design, emphasizing that increasing the number of elements can significantly enhance system performance while meeting the cost and power consumption requirements of future networks.
\begin{figure}
	\label{PAAbefore}
	\centering
	\subfigure[System sum-rate for DL.]{
		\begin{minipage}{7.45cm} 
			\includegraphics[width=\textwidth]{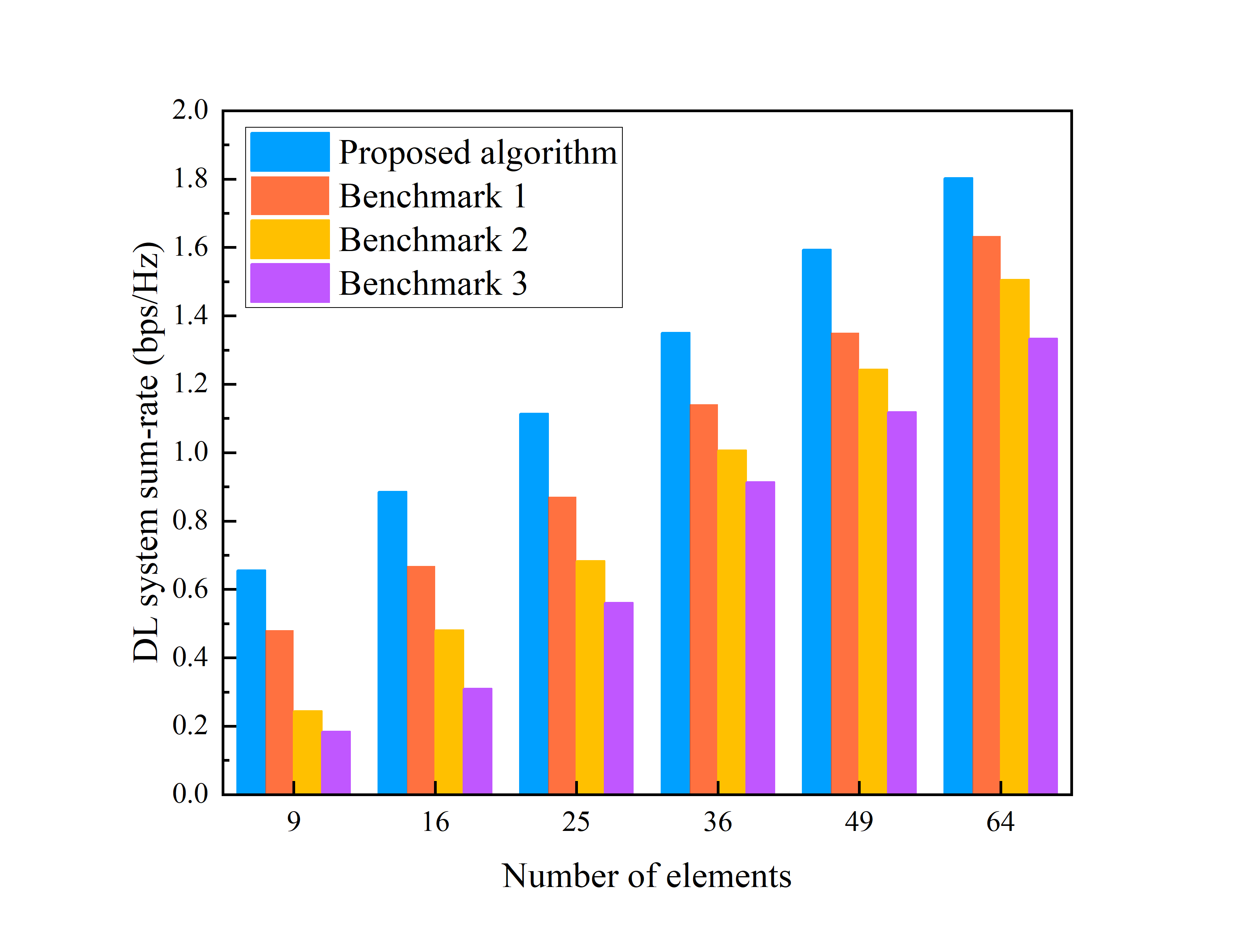} \\
		\end{minipage}
	}
	\subfigure[System sum-rate for UL.]{
		\begin{minipage}{7.45cm}
			\includegraphics[width=\textwidth]{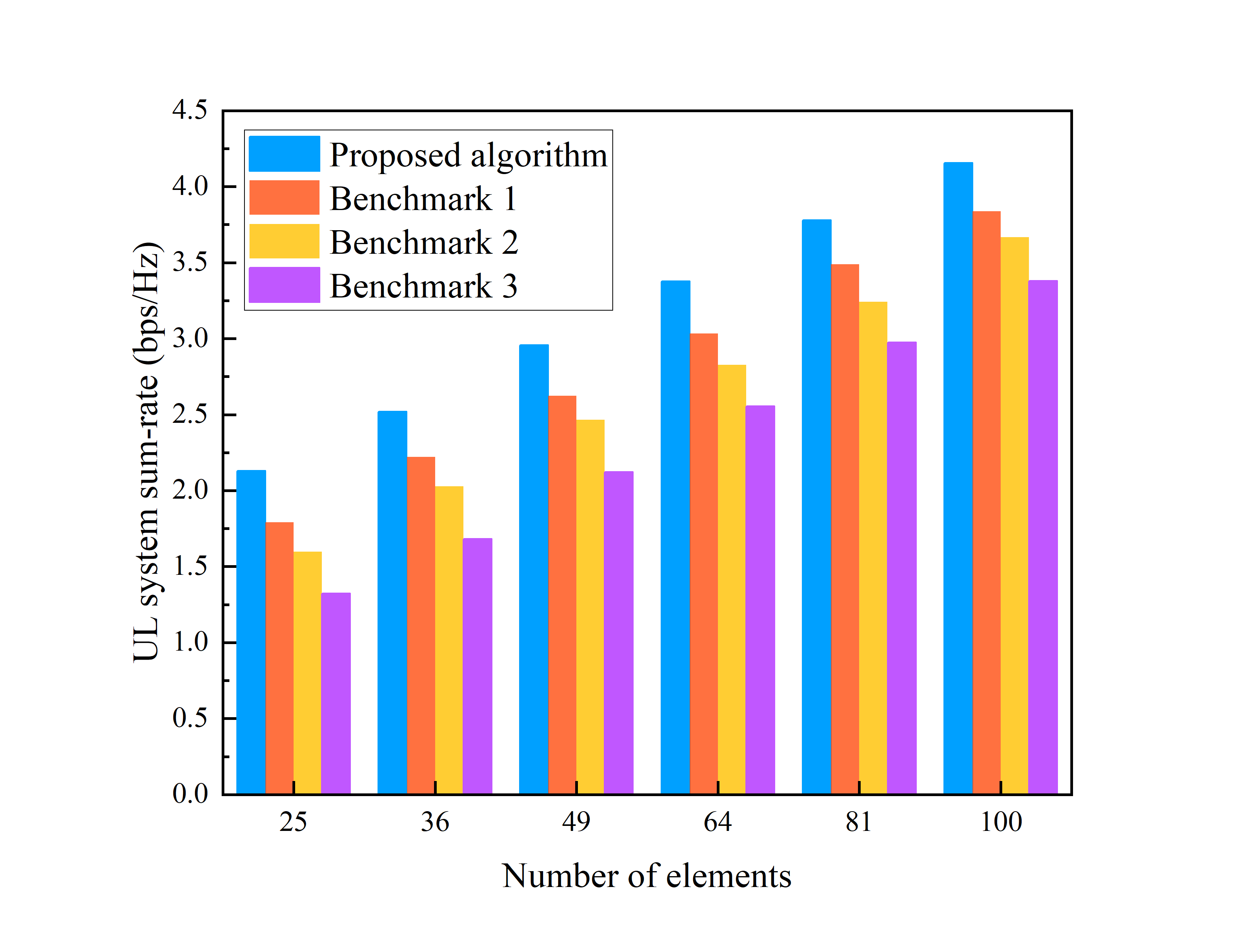} \\
			
		\end{minipage}
	}
	\caption{System sum-rate varies with the number of RIS transmissive elements.} 
\end{figure}

\section{Future of TRTC: Exploring Potential Directions and Opportunities}
Regarding the advantages of the proposed TRTC architecture, there exist two potential research directions and opportunities for future investigations. 
\subsection{NOMA Design in TRTC}
For UL communications in TRTC, the horn feed employs a single antenna. Consequently, alongside the traditional orthogonal multiple access (OMA) scheme, NOMA is a highly promising access scheme. Similarly, for DL, NOMA can enable massive user access in future networks. In NOMA, it puts forward new requirements for the optimization of RIS. RIS needs to dynamically adjust the coefficient according to the power levels of different users to maximize the overall performance of the system. This increases the complexity of the optimization problem. Meanwhile, the power superposition principle may lead to increased interference, which affects the stability and reliability of the system. In addition, optimizing RIS coefficients to improve the accuracy of SIC requires precise channel state information and complex algorithm design, which may increase the computational burden of the system. Consequently, designing a TRTC system that takes into account both the SIC decoding order and the transmissive coefficient is a novel and important undertaking.
\subsection{Artificial Intelligence (AI)-Based Desgin in TRTC}
The current optimization-based design for DL and UL communication may not be suitable for highly dynamic scenarios and responsive network environments. AI-based design methods can potentially address the limitations of traditional optimization methods in solving problems. Despite the limited number of RIS components, AI can still play an important role in the optimization process. AI can improve system performance by learning complex channel characteristics and user requirements to achieve more efficient RIS optimization algorithms. AI technology can quickly adapt to changes in dynamic environments and provide real-time decision support for RIS optimization, especially in complex scenarios with multiple users and multiple channels. Utilizing AI for CSI acquisition in DL and UL, as well as for adaptive configuration of system parameters through pre-training, could accelerate the implementation of TRTCs in future communications. However, the development of AI models for this framework poses interesting and challenging problems that require further exploration.
\section{Conclusions}
This paper presents an overview of an efficient TRTC architecture, with the primary objective of attaining a low power consumption and cost-effective transceiver for B5G and 6G networks. Leveraging the RIS to adjust its transmissive coefficient facilitates DL beamforming and UL signal enhancement. Since the TRTC architecture has the characteristics of scalability, flexibility, adaptability, and environmental protection, it can meet the diverse needs of future networks. Given that the design and optimization of the TRTC are still nascent and not fully explored, it is anticipated that this paper will provide valuable insights and guidance for future theoretical research and technical design. Moreover, the integration of low-cost and low-power RIS is expected to encompass the entire closed-loop wireless transmission link, comprising transmitters, auxiliary wireless communications, and receivers in future networks. The execution of a joint optimization design for the RIS-based communication system is anticipated to yield optimal overall performance.

\bibliographystyle{IEEEtran}
\bibliography{reference}
\begin{IEEEbiographynophoto}{Zhendong Li}
(lizhendong@xjtu.edu.cn) is an Assistant Professor with the School of Information and Communication Engineering, Xi'an Jiaotong University, Xi'an, China. His research interests include reconfigurable intelligent surface (RIS), unmanned aerial vehicle (UAV) communications, space-air-ground (SAG) networks, the Internet-of-Things (IoT), and wireless resource management in future wireless networks.
\end{IEEEbiographynophoto}
\begin{IEEEbiographynophoto}{Wen Chen}
(wenchen@sjtu.edu.cn) is a tenured Professor with the Department of Electronic Engineering, Shanghai Jiao Tong University, China, where he is the director of Broadband Access Network Laboratory. His research interests include multiple access, wireless AI and reconfigurable intelligent surface enabled communications. He has published more than 180 papers in IEEE journals with citations more than 10,000 in Google scholar.
\end{IEEEbiographynophoto}
\begin{IEEEbiographynophoto}{Qingqing Wu}
(qingqingwu@sjtu.edu.cn) is an Associate Professor with Shanghai Jiao Tong University. His current research interest includes intelligent reflecting surface (IRS), unmanned aerial vehicle (UAV) communications, and MIMO transceiver design. He has coauthored more than 100 IEEE journal papers with 30 ESI highly cited papers and 9 ESI hot papers, which have received more than 26,000 Google citations. He was listed as the Clarivate ESI Highly Cited Researcher since 2021.
\end{IEEEbiographynophoto}
\begin{IEEEbiographynophoto}{Ziwei Liu}
(ziweiliu@sjtu.edu.cn) is currently pursuing the Ph.D. degree with the Broadband Access Network Laboratory, Department of Electronic Engineering, Shanghai Jiao Tong University (SJTU), Shanghai, China. His research interests include reconfigurable intelligent surface (RIS), rate splitting multiple access (RSMA), and wireless resource management in future wireless networks.
\end{IEEEbiographynophoto}

\begin{IEEEbiographynophoto}{Chong He}
(hechong@sjtu.edu.cn) is an Associate Professor with Shanghai Jiao Tong University. Dr. He has authored and coauthored over 90 articles in IEEE journals and conferences. His research interests include phased arrays, reconfigurable intelligent surface, DOA estimation, wireless location and multiple access wireless communications.
\end{IEEEbiographynophoto}
\begin{IEEEbiographynophoto}{Xudong Bai}
(baixudong@nwpu.edu.cn) is an Associate Professor with the School of Microelectronics, Northwestern Polytechnical University (NWPU), China. He has authored or coauthored more than 80 papers and holds more than 10 patents in antenna and metamaterial technologies. His current research interests include electromagnetic metasurfaces, low-cost phased arrays, and OAM-EM wave propagation and antenna design.
\end{IEEEbiographynophoto}
\begin{IEEEbiographynophoto}{Jun Li}
(jleesr80@gmail.com) is a Professor at the School of Information Science and Engineering, Southeast University, Nanjing, China. He was a visiting professor at Princeton University from 2018 to 2019. His research interests include network information theory, game theory, distributed intelligence, multiple agent reinforcement learning, and their applications in ultra-dense wireless networks, mobile edge computing, network privacy and security, and industrial Internet of things.
\end{IEEEbiographynophoto}
%
%
%
%
%
%
%
%

\end{document}